\documentstyle[epsfig,aps]{revtex}

\newcommand{\aleq}{\mbox{\ \raisebox{-.9ex}{$\stackrel{\textstyle <}{\sim}$}\ }}
\newcommand{\ageq}{\mbox{\ \raisebox{-.9ex}{$\stackrel{\textstyle
>}{\sim}$}\ }}

\begin{document}

\title{Electrostatic Edge Instability of Lipid Membranes}

\author{M.   D.  Betterton$^1$ and Michael P. Brenner$^2$}
\address{$^1$ Department of Physics, Harvard University, Cambridge, MA 02138 \\
$^2$ Department of Mathematics, MIT, Cambridge, MA 02139}

\maketitle

\begin{abstract}
This paper considers the effect of electrostatics on the stability of
a charged membrane.  We show that at low ionic strength and high
surface charge density, repulsion between charges on the membrane
renders it unstable to the formation of holes.  A free straight edge
is unstable to modulations with wavelength longer than the Debye
screening length.  Hence at low ionic strength, membranes will
disintegrate into vesicles.  We use these results to interpret
stable holes in red blood cell ghosts (Steck {\it et. al.}, {\sl
Science} \ 168:255(1970)).

\end{abstract}

Lipid membranes, the sheets which form the boundaries of cells, must
maintain their integrity for cells to live\cite{albert}.  Since a
tear in a membrane exposes the hydrophobic interior of the sheet
to water, tradition maintains that membranes have 
a high line tension (energy per unit length of exposed edge) and
never spontaneously form holes.  We show that membranes
can form holes in a simple model of a
{\sl charged} membrane. By examining the competition between
electrostatic repulsion and line tension, we find a parameter range where
electrostatics dominates and membranes are unstable to hole
formation.  Although much previous work examines 
electrostatic effects on lipid assemblies \cite{israel} and the bending
moduli of membranes \cite{win88,Siggia-bend,dup90}, we are only aware
of one study addressing electrostatics and line tension \cite{pet80}.

\begin{figure}[p]
\centerline{\epsfysize=2.5in\epsfbox{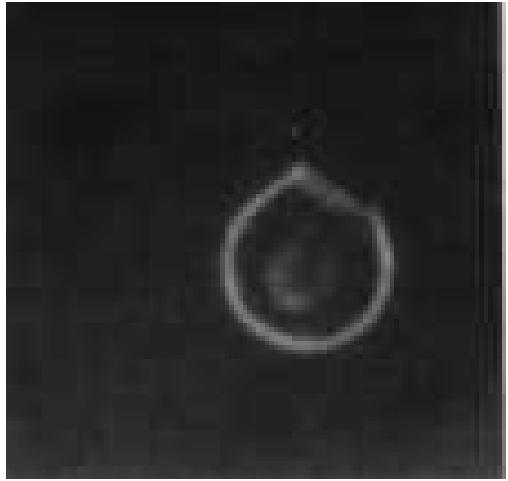}}
\caption[]
{Image of hole in red blood cell ghost, from
Lieber and Steck \protect\cite{Steck1} Figure 7. }
\end{figure}

These calculations were motivated by experiments which observed
stable holes in red blood cell ghosts
\cite{ste70,Steck1,Lew-Vesicle,lew82}, the size of which depends on
the ionic strength of the surrounding fluid
\cite{Steck1} (Fig. 1). We compare our calculation with these
experiments, and argue that the observed morphologies
require the spectrin skeleton.

Consider a thin, axisymmetric membrane with constant charge density
$\sigma$ on each side (Fig. 2a).  The membrane attracts ions of the
opposite sign, creating a screening layer (of thickness $\kappa^{-1}$
in Debye Huckel theory).  We assume that the membrane is flat; this is
a good approximation when the hole radius $R$ and screening length
$\kappa^{-1}$ are small compared to the membrane radius of curvature.
The plane of the membrane is at
$z=0$. Throughout this paper, the membrane area and charge are
held constant.
\begin{figure}[p]
\centerline{\epsfysize=2in\epsfbox{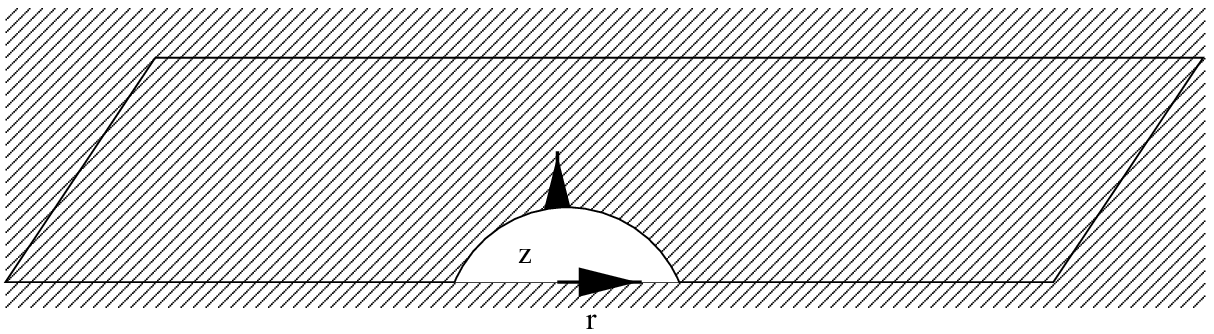}}
\centerline{\epsfysize=2in\epsfbox{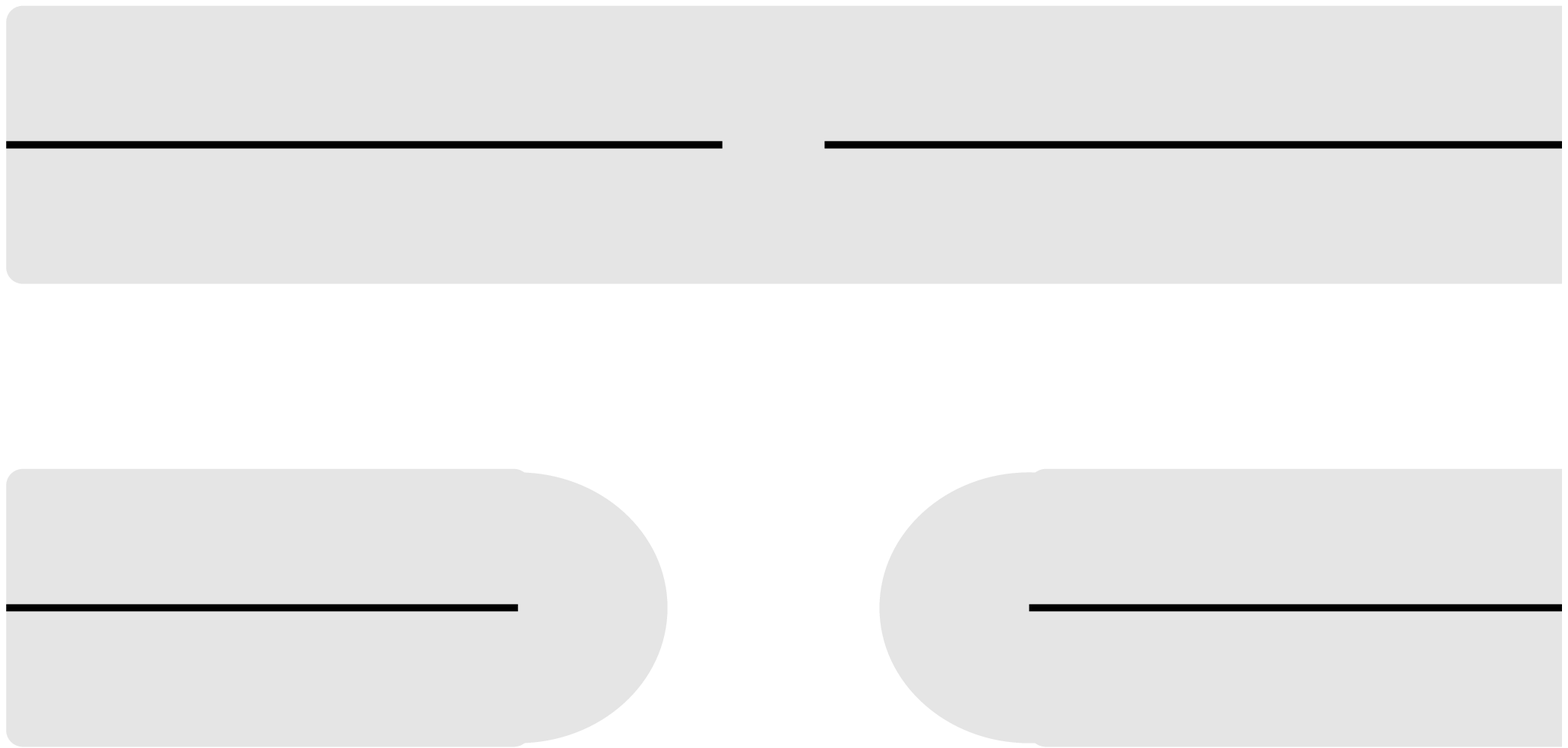}}
\caption
{ (a) Sketch of a membrane hole;  shading denotes constant charge
density. (b) Sketch of the screening cloud (shaded)
around the membrane (bold line).  When $R< \kappa^{-1}$ (top),
the screening cloud fills the hole.  When $R>\kappa^{-1}$(bottom), the
 cloud extends only $\kappa^{-1}$ from the edge.  }
\label{sketch}
\end{figure}

Repulsion between charges on the membrane favors the creation and
expansion of holes.  Heuristically, the electrostatic energy is
determined by the volume of the screening cloud:  the total
charge in the screening cloud is constant, so expanding the cloud
lowers its energy.  First, when the
hole is small relative to a screening length (Fig. 2b), the cloud expands
by a volume $\sim R^2 \kappa^{-1}$ over the hole.  Since the
electric field strength is $\sim \sigma\epsilon_w^{-1}$ ($\epsilon_w$
is the dielectric constant of water), the energy decrease is 
$\sim \sigma^2 \kappa^{-1} \epsilon_w^{-1}R^2$. The energy increases
because an edge of membrane is exposed; if the
membrane has a line tension $\gamma$, the energy cost of a small hole
is $ U \sim \gamma R- \sigma^2
\kappa^{-1}\epsilon_w^{-1}R^2 $. 
For small $R$  line tension dominates, and a hole 
closes. 

In the opposite limit, the hole is much larger than a screening
length.   The cloud does not fill the hole, but
inhabits
an additional volume equal to 
that of a tube of radius $\kappa^{-1}$
and length $R$.  
The total energy is 
$U \sim ( \gamma-\sigma^2\kappa^{-2}\epsilon_w^{-1}) R$. 
Depending on $\gamma$, $\sigma$, and
$\kappa$, the  energy
can be positive {\em or negative}.  The latter case corresponds
to instability and hole growth.

We quantify this argument  by  calculating the energy
of an idealized membrane hole.  The relevant dimensionless parameter
$P$ is the ratio of the line tension to the electrostatic energy per
screening length of edge $ P = \epsilon_w \kappa^2
\gamma/\sigma^2 $. We
 show that when line tension dominates, holes
close. When $P \ll 1$, large holes grow.  At
intermediate $P \approx 2$ metastable holes exist due to the
competition between energies.  We then discuss the stability of a
straight membrane edge to sinusoidal perturbations with wavenumber
$q$.  We show that at small $P$, and in the long wavelength limit $q
\ll \kappa $, small perturbations lower the overall energy.  In this
regime, a flat membrane with a free edge will generically break into
pieces of minimum size $O(\kappa^{-1})$.  The final section of the paper relates
these results to the ghost experiments.

To compute the electrostatic energy of a hole, we use the linearized
Poisson-Boltzmann equation \cite{linear} $\nabla^2 \phi = \kappa^2
\phi$. The boundary conditions are $\phi\rightarrow 0$ as $ z
\rightarrow \infty $, $\partial_z \phi(z = 0) = -2 \pi
\sigma/\epsilon_{w}, r > R$ and $\partial_z \phi(z = 0) = 0$ for $r <
R$.  The latter comes from symmetry of the potential about $z=0$.
We Hankel transform to find
\begin{eqnarray}
 \phi (r, z) &=&
	 \frac{ 2 \pi \sigma  e^{- \kappa z} }{\epsilon_w \kappa} \\
	&-& \frac{2 \pi \sigma R}{\epsilon_w} \int_{0}^{\infty} \frac{
	dk\: J_0 (kr)\: 
	J_1(kR)}{ \sqrt{ k^2   + \kappa^2 } } 
	 e^{-z \sqrt{k^2   + \kappa^2 } } \nonumber
\end{eqnarray}
where  $ J_0$ and $ J_1$  are Bessel functions. Without a
hole ($R = 0$), the potential reduces 
to that of an infinite charged sheet. 

The electrostatic energy is
$U = 2 \pi \int_{0}^{\infty}  r dr \: \sigma  \phi(z = 0)$; we assume the total
charge and the membrane area remain constant.
 The electrostatic contribution competes with
the energy from line tension, $2 \pi
\gamma R$.
 The energy change, nondimensionalized
by $ 2 \pi \sigma^2/(\epsilon_w\kappa^3)$, 
 depends
on the parameter $ P =\epsilon_w \kappa^2  \gamma / \sigma^2 $ and the
dimensionless radius $\xi = \kappa R $
\begin{equation}
 \Delta U = P  \xi -\pi  \xi^2  +2 \pi  \xi ^3 \int_{0}^{  \infty } \frac{ dx
	J_1(x)^2 }{ x \sqrt{x^2  +  \xi^2 } }
\end{equation}
When the hole is small ($ \xi  \ll 1 $), the integral in the third
term is approximately $4/(3\pi)$,
so the energy change of a hole is
$
\Delta U \approx P  \xi -\pi  \xi^2  + 8/3 \xi ^3.
$ Thus, for small holes the line tension
dominates over electrostatic effects.  For larger   $\xi$, the
energy is shown in Fig. 3. When $P\ageq 2.2$, the energy increases
monotonically as a function of $ \xi $, so holes always close.  For
$2.0$ \mbox{\raisebox{-.9ex}{$\stackrel{\textstyle <}{\sim}$}}
$P \aleq 2.2$, we find a local energy
minimum with $\xi \sim 1$. In principle, holes in a membrane could be
observed in this region of metastability. In practice, however, we
believe that the range of parameters in which such holes exist
is prohibitively narrow.  When $P
\aleq 2$, the energy decreases linearly at large $\xi$, so that large
holes will grow.  In this regime there is an activation energy barrier
to hole formation of size $\epsilon_w \kappa \gamma^2
\sigma^{-2}/2$.

\begin{figure}[p]
\centerline{\epsfysize=2.75in\epsfbox{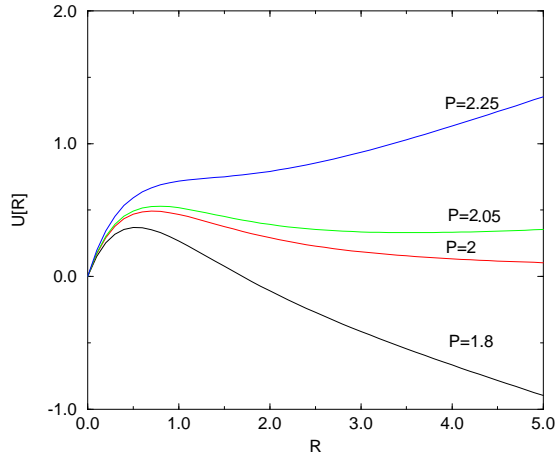}}
\caption{
Energy of a hole as a function of dimensionless radius $\kappa R$, for
$P=1.8, 2, 2.05$ and $2.25$. A typical barrier height is 10 kT (for
$P=1$, which corresponds to $n$=1 mM, 
$\gamma=10^{-6}$ erg, and  $\sigma= 0.2 \: e/\mbox{nm}^2 $).  }
\label{enerfig}
\end{figure}

A membrane with a growing hole ultimately has a long edge; is this
edge stable? We consider a straight edge
segment (at $x=0$) modulated by a sinusoidal
perturbation $x=\epsilon {\rm cos}(q y)$.  For this problem the charge
density is $\sigma=\sigma_0 \Theta( x +\epsilon \cos qy ),$ where
$\Theta$ is a step function.  The energy  follows from the
linearized Poisson-Boltzmann approximation by expanding  the
potential and the charge density in powers of $\epsilon$.  The first
nonzero correction to the energy from the edge perturbation occurs at
$O(\epsilon^2)$, and this change in the electrostatic energy (in
dimensionless units) due to the edge modulation is $-\epsilon^2
\kappa^3 q^{-1} {\rm ln}(1+\kappa^{-2}q^2)$.


The perturbation increases the length of edge by $\Delta L= \pi/2
\epsilon^2 q$, resulting in an energy increase $\gamma\Delta L$ from
line tension.  
For large $q$ line tension dominates,  
increasing the energy.
For small  $q \ll \kappa$, the total energy change
is
\begin{equation}
\Delta U = \frac{\epsilon^2 q \kappa}{2} \biggl( \frac{P}{2}-1 \biggr) + O(q^2),
\end{equation}  
If (a) electrostatic effects dominate ($P<2$) and (b) the perturbation
wavelength is greater than the screening length, modulations of the
edge grow.  The growth continues as long as the circumference of
membrane fragments is longer than the screening length.
When the fragment radius is of order $2 \pi
\kappa^{-1}$ the instability will stop,and the fragment may close  to
form a vesicle.  In the
absence of charge, the fragment size must be $\ageq 10$ nm \cite{bend}, so the
energy associated with the free edge is larger than the bending energy required to
close into a vesicle.  With electrostatic effects
included, it seems additional forces are necessary to promote vesicle
formation.  A candidate for red
blood cells is spontaneous curvature: the higher charge density on the
inner face of the bilayer \cite{Steck1} means the
membrane prefers to bend towards the outer face\cite{Siggia-bend}.  
Understanding  the final vesicle sizes requires analyzing both the
 electrostatic edge instability and spontaneous
curvature---in the Lew experiments,
\cite{Lew-Vesicle} spontaneous curvature effects play an important
role in the experimental phenomenology.

Observations of stable holes in red blood cell ghosts motivated this
work\cite{holes}.  Human red  cells, if burst by  osmotic
stress and placed in a low-salt buffer, break into
vesicles\cite{ste70}, which  typically
show the cytoplasmic (inner) side of the membrane facing outwards.
Studies of this process
\cite{Steck1,Lew-Vesicle} illustrate that vesiculation is preceded by
the formation of {\em stable} holes in the membrane, whose size
depends  on the salt concentration of the solution and the
membrane charge density.  Lieber and Steck
\cite{Steck1} showed that (a) the hole size increases with decreasing
salt concentration; and (b) a decrease (increase) in the effective
membrane charge density---via charged intercalators inserted into the
membrane---decreases (increases) the hole size.  Our model qualitatively explains these trends,
but uncertainties in our model and in experimentally measured
parameters prevent precise numerical calculation.  We have
neglected (a) differences in charge density on the
two sides of the membrane, (b) the insulating interior of the
bilayer\cite{Siggia-bend}, and (c) nonlinear effects.   Although
there are experimental indications that some of these
effects might be relevant\cite{calcium}, in the absence
of quantitative information for
the line tension
and charge density, the simple model presented here is adequate.

Fig. 4 shows the stability
boundaries as a function of the surface charge density $\sigma$ and
the salt concentration $n$.  We assume $\gamma=10^{-6}$ dyn,
consistent with recent experiments\cite{zhe93}.  The phase boundaries
follow $\sigma \propto n^{1/2}$. For typical membrane
charge density $\sigma \sim 0.2 \: e/ \mbox{nm}^2 $, the crossover
point $P \sim 2$ occurs at a salt concentration of 2 mM, which is in the
range of the Lieber-Steck experiments.  When the energy
barrier for hole formation is of order a thermal energy $k_BT$,
thermal fluctuations produce holes in the membrane.  The
dashed line in Fig. 4 shows the borderline for this
instability\cite{nonlinear}.  Since the barrier scales 
like $\gamma^2$,  the exact position of this important
borderline is sensitive to the line tension $\gamma$.

\begin{figure}[p]
\centerline{\epsfysize=2.75in\epsfbox{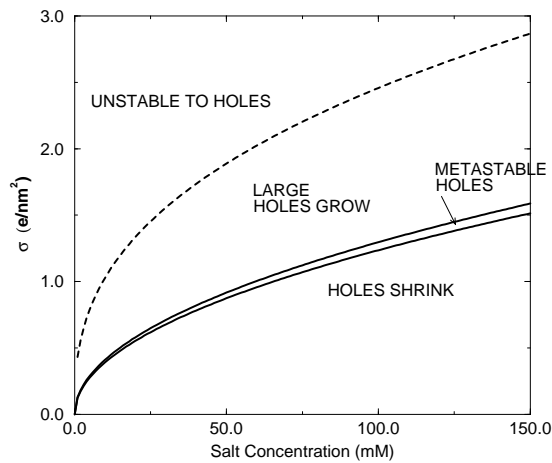}}
\caption{
Phase Diagram, depicting regimes of shrinking, growing and
metastable holes, with $\gamma=10^{-6}$ dyn. The upper and lower solid
lines depict $P=2$ and $2.2$. The electrostatic edge
instability is present everywhere above the solid lines. Note that the
salt concentration (assuming a 1:1 salt such as NaCl) is related to
the Debye length via $\kappa^2 = n/\ell^2$, where $\ell^2 = \epsilon_w
k_B T/(8\pi e^2 N)$, with $N$ Avogadro's number.  }
\label{phasediag}
\end{figure}

The electrostatic model qualitatively explains the dependence on salt
concentration and membrane charge density in the experiments.
However, electrostatic effects  are not sufficient: the edge
instability predicts that holes either grow indefinitely (resulting in
vesiculation) or close completely.  The observations
\cite{Steck1,Lew-Vesicle} of stable holes for long time periods
contradict this prediction and indicate that a stabilizing effect is
necessary.

\begin{figure}[p]
\centerline{\epsfysize=2in\epsfbox{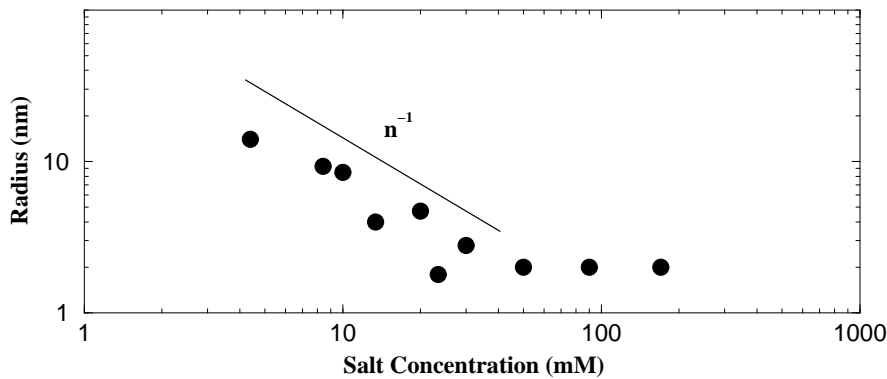}}
\caption[]
{
Radius of hole versus salt concentration	
from Ref. \protect\cite{Steck1}.  The solid line is $R\sim n^{-1}$. }
\label{steck_fig}
\end{figure}

Several aspects of the experiments indicate that the spectrin
skeleton---a protein mesh anchored to the
membrane---is this stabilizing element. The electron micrographs of
Lew {\it et. al.}
\cite{Lew-Vesicle} show that the spectrin network is intact  before
 vesiculation occurs, and as the ghosts disintegrate into
vesicles the spectrin mesh detaches from the membrane.  Lieber and
Steck \cite{Steck1} found that proteins known to covalently cross-link
spectrin stabilize the hole, preventing changes in radius as
the ionic strength of the solution is changed.  Conditions
known to promote breakdown of  spectrin (such as increase in
temperature or digestion by enzymes\cite{Steck1}) cause the ghosts to
vesiculate.  If the spectrin
provides the restoring force needed to stabilize the membrane, force
balance implies $k_s R = -dU/dr$, where $k_{s}$ is the spectrin area
expansion modulus and $U$ the hole energy computed above.  At small
$P$, we find
\begin{equation}
R = \frac{4 \pi \sigma^2 \ell^2}{\epsilon_w k_s} \frac{1}{n}.
\end{equation}
Figure \ref{steck_fig} shows
Lieber and Steck's
data \cite{Steck1}, replotted
on a double logarithmic scale.
The data
obey this law over a decade in hole size.  
To check the consistency we fit the prefactor of the scaling law.  A best
fit gives
the prefactor $c= 7.0 \cdot 10^{-6}$ cm M.   For 
$\sigma$  in units of
$e/\mbox{nm}^2$ and $k_s$ in dyn,
we find $\sigma^2/k_s = 0.02$. This ratio is consistent
with the charge density $\sigma= 10^{-2}$ and bulk
modulus $k_s= 10^{-2}$ from
independent
measurements \cite{spectrin}.

The mechanism of vesiculation by electrostatic edge instability does
not require spontaneous curvature; hence it differs from other
explanations of vesiculation\cite{hau89,win88,Siggia-bend}. There are
conditions under which vesiculation appears to be driven by
spontaneous curvature \cite{li86}.  It seems likely that both
mechanisms for vesiculation operate in practice.

For physiological conditions, membrane charge densities range
from $-0.03$ to $-0.24 \ |e|/\mbox{nm}^2$ 
\cite{Siggia-bend,cho80,sac83} and NaCl concentrations are
around $150$mM.  The phase diagram Fig. 4 predicts that such membranes
should be stable, although conditions {\em in vivo} are within an
order of magnitude of the stability boundary.  This indicates that the
edge instability could be important in living systems.  Partition of
the nuclear envelope (the membrane surrounding the nucleus) during
cell division provides a possible example \cite{war93,roos}. Because
it is a double bilayer\cite{albert}, its effective line tension is
much smaller than a typical membrane. The qualitative behavior of the
nuclear envelope is similar to the phenomenology described here:
during most of the cell cycle, the nuclear envelope is attached to a
protein meshwork called the nuclear lamina\cite{war93}. Near the start
of cell division, the lamina disassembles---and the nuclear envelope
is observed to fill with holes and vesiculate\cite{roos}. The many
vesicles are divided between the two daughter cells during division.

{\bf Acknowledgements:}  We are grateful to Professor Ted Steck, for
patiently
introducing us to his experiments, and
David Lubensky for useful discussions.
MPB acknowledges support from the A.P. Sloan foundation.
MDB
acknowledges support by a Fellowship from the Program in Mathematics
and Molecular Biology at the Florida State University, with funding
from the Burroughs Wellcome Fund Interfaces Program.


\begin{thebibliography}{10}

\bibitem{albert}
B. Albert {\it et~al.}, {\em Molecular Biology of the Cell} (Garland
  Publishing, Inc., New York, 1989).

\bibitem{israel}
J. Israelachvili, D. Mitchell, and B. Ninham, J. Chem. Faraday Trans.
2 {\bf 72}, 1525 (1976) 


\bibitem{win88}
M. Winterhalter and W. Helfrich, J. Phys. Chem. {\bf 92},  6865  (1988).
M. Winterhalter and W. Helfrich, J. Phys. Chem. {\bf 92},  327  (1992).

\bibitem{Siggia-bend}
T. Chou, M. Jaric, and E.~D. Siggia, Biophys. J. {\bf 72},  2042
  (1997).

\bibitem{dup90}
B. Duplantier et. al.
Phys. Rev.
  Lett. {\bf 65},  508  (1990).

\bibitem{pet80}
A.G. Petrov et. al.
Adv. Liq. Cryst. Research (L. Bata,
ed), {\bf 2} 695 (1980).  This paper briefly discusses
line tension in an external electric field.

\bibitem{ste70}
T. Steck et. al.
Science {\bf 168},
  255  (1970); T. Steck and J. Kant, Meth. Enzym. {\bf 31},
  172(1974).

\bibitem{Steck1}
M.~R. Lieber and T.~L. Steck, J. Bio. Chem {\bf 57},  11651  (1982).
M.~R. Lieber and T.~L. Steck, J. Bio. Chem {\bf 57},  11660  (1982).

\bibitem{Lew-Vesicle}
V. Lew et. al.
J. Cell
  Biol. {\bf 106},  1893  (1988).

\bibitem{lew82}
V.~L. Lew, S. Muallem, and C.~A. Seymour, Nature {\bf 296},  742  (1982).

\bibitem{linear} The linearized approximation is valid
whenever $e\phi\ll k_B T$.
 For physiological salt concentrations, this condition holds, but a
 crossover occurs at lower salt concentration; $e\phi \sim k_B T$
 for 25 mM salt.

\bibitem{calcium}  An example of a
nontrivial electrostatic effect observed
in the experiments \cite{Steck1} is that
varying the concentration of divalent cations has a much larger effect
than would be anticipated from changing the Debye length.

\bibitem{zhe93}
D.~V. Zhelev and D. Needham, Biochemica et Biophysica Acta {\bf 1147},  89
  (1993).

\bibitem{nonlinear}
The dashed curves in Fig. 4 occur outside the regime
where linearized Poisson-Boltzmann is valid. The
exact solution of the Poisson-Boltzmann equation for an infinite
plane indicates that nonlinearity leads to stronger
confinement of the screening cloud
(effectively decreasing the screening length).
Nonlinearities do {\it not} change the scaling of the electrostatic
energy with the hole radius $R$, and therefore do not qualitatively
affect the results.

\bibitem{bend} 
W. Helfrich, Phys. Lett. {\bf 50A}, 115 (1974); 
P. Fromherz, Chem. Phys. Lett. {\bf 94}, 259 (1983).

\bibitem{holes}
More recent works focus on holes in membranes
for other reasons. See \cite{zhe93};
J. Moroz and P. Nelson, Biophys. J. {\bf 72}, 2211-16   (1997);
R. Bar-Ziv et. al.
Phys. Rev. Letter {\bf 75},  3481  (1995).
W. Sung and P.J. Park, Biophys. J. {\bf 73} 1797 (1997).

\bibitem{spectrin}
H. Engelhardt and E. Sackmann, Biophys. J. {\bf 54}, 495 (1988);
R. Waugh and E. Evans, Biophys. J. {\bf 26}, 115(1979); D. Discher
et. al. 
Science {\bf 266},1032 (1994); Y. Kantor and D. R. Nelson, Phys. Rev.
A.{\bf 36} 4020(1987).

\bibitem{hau89}
H. Hauser, Proc. Nat. Acad.Sci {\bf 86},  5251  (1989).

\bibitem{li86}
W. Li and T.~H. Haines, Biochemistry {\bf 25},  7477  (1986).

\bibitem{cho80}
W. Chow and J. Barber, J. Biochem. Biophys. Methods {\bf 3},  173  (1980).

\bibitem{sac83}
F. Sack, D.Priestley, and A. Leopold, Planta {\bf 157}, 511 (1983).

\bibitem{war93}
G. Warren, Ann. Rev. Biochem.  323  (1993).

\bibitem{roos}
U. Roos, Chromosoma {\bf 40}, 43 (1973).

\end{thebibliography}
\end{document}